\documentclass{book}
\newcommand{\be}{\begin{equation}}
\newcommand{\ee}{\end{equation}}
\begin{document}

\title{Evaluation of Casimir energies through spectral
functions}


\author{E.M. Santangelo \\Departamento de F\'{\i}sica, Facultad de Ciencias Exactas,
\\Universidad Nacional de La Plata \\ C.C. 67, 1900 La Plata,
Argentina\\email: mariel@obelix.fisica.unlp.edu.ar}

\date{St. Petersburg, December 2000}

\maketitle

\section*{Abstract}

This is an introductory set of lectures on elliptic differential
operators and boundary value problems, and their associated spectral
functions. The role of zeta functions and traces of heat kernels in
the regularization of Casimir energies is emphasized, and the
renormalization issue is discussed through simple examples.

\tableofcontents

\chapter{Zero point energy in field quantization}

In this chapter, the concept of Casimir energy is introduced. Its
evaluation in a very simple example is performed, through two
different regularization methods. Both results are shown to be
consistent after renormalization.\footnote{Suggested bibliography:
\cite{itzykson,casimir,plunien,moste}}

\section{Free massive neutral scalar field in the whole
Minkowski space-time}

In order to introduce the problem of vacuum (ground state) energies
in Quantum Field Theory, we start by briefly reviewing the simple
example of a free massive neutral scalar field in the whole
Minkowski space-time.

The classical Lagrangian density is, in this case, given by
\[
{\cal L}=\frac12 \left(\partial \phi\right)^2 -\frac12 m^2
\phi^2\,,
\]
where $m$ is the mass of the field. Through Euler- Lagrange, the
classical Klein-Gordon equation of motion follows
\[
\left({\partial}^{\mu}{\partial}_{\mu}+ m^2 \right)\phi =0\,.
\]

The momentum conjugate to the field is here given by
\[
\pi =\frac{\partial \cal{L}}{\partial (\partial_0
\phi)}=\partial_0\phi\,,
\]
so that the classical Hamiltonian turns out to be
\[
H=\int\,d^3 x \frac12\left[\pi^2 +(\,\,\, \phi)^2 +m^2 \phi^2
\right]\,.
\]

The transition to the quantum theory is achieved by rising $\phi$
and $\pi$ to the condition of operators, transforming Poisson
brackets into commutators and imposing, at equal times,
\begin{equation}
\left[\hat{\phi}(t,\vec{x}),\hat{\pi}(t,\vec{x'})\right]=i
{\delta}^3 (\vec{x}-\vec{x'})\,, \label{qcond}
\end{equation}
with the other commutators vanishing.

Now, going to the momentum representation, with $k^{\mu}
=(k^0,\vec{k})=(\omega_{\vec{k}},\vec{k})$ and
$\omega_{\vec{k}}=\omega_{-\vec{k}}=\sqrt{(\vec{k}^2 +m^2)}>0$, both
fields can be expanded as
\[
\hat{\phi}(t,\vec{x})=\int\,d\tilde{k}\left[\hat{a}(\vec{k})
e^{-ik.x} +\hat{a}^{\dag}(\vec{k}) e^{ik.x}\right]\]
\[
\hat{\pi}(t,\vec{x})=-i\int\,d\tilde{k}\left[\hat{a}(\vec{k})
e^{-ik.x} -\hat{a}^{\dag}(\vec{k}) e^{ik.x}\right]=\partial_0
\hat\phi (t,\vec{x})\,,
\]
where $d\tilde{k}=\frac{d^3 k}{(2\pi)^3 2{\omega}_{\vec{k}}}$ is the
Lorentz invariant measure. Notice, in passing, that
${\omega}_{\vec{k}}$ are the square roots of the eigenvalues of the
operator $\Delta+m^2$.

The quantization conditions (\ref{qcond}) then translate into
\[
[\hat{a}(\vec{k}),\hat{a}^{\dag}(\vec{k'})]=(2\pi)^3 2\omega_k
\delta^3 (\vec{k}-\vec{k'})\,,
\]
with all other commutators vanishing.

As to the Hamiltonian operator, it takes the form
\[
\hat{H}=\frac12 \int
d\tilde{k}\,\omega_{\vec{k}}[\hat{a}^{\dag}(\vec{k})\hat{a}(\vec{k})+\hat{a}(\vec{k})
\hat{a}^{\dag}(\vec{k})]=
\]
\[
\frac12 \int
d\tilde{k}\,\omega_{\vec{k}}[2\hat{a}^{\dag}(\vec{k})\hat{a}(\vec{k})+
(2\pi)^3 2\omega_{\vec{k}}"{\delta}^{3}(0)"]\,. \}\]

Now, the vacuum state is defined through
\[
\hat{a}(\vec{k})|0>=0\,\,<0|0>=1\,,
\]
and
\[
<0|\hat{H}|0>=\frac12 \int d^3 k
\,\omega_{\vec{k}}``{\delta}^{3}(0)"\,,
\]

The factor $``{\delta}^{3}(0)"$ is meaningless, and comes from the
infinite size of the system. It can be given a sense, for instance,
by enclosing the system in a cubic box of volume $V$, imposing
periodic boundary conditions on the field, and then taking the
infinite volume limit.

When doing so, one has
\[
<0|\hat{H}|0>=\frac{V}{2} \int \frac{d^3 k}{(2\pi)^3}
\sqrt{(\vec{k}^2+m^2)}\,,
\]
with $V$ the (infinite) space volume.

This shows how a divergent zero-point energy per unit volume arises
in the case of a free scalar field: The divergence is due to the sum
of the zero-point energies of an infinite number of oscillators. For
the free theory in the whole Minkowski space, one defines this vacuum
expectation value to be zero, by imposing the normal ordering
prescription. But, then, a question naturally arises: What's the
value of the vacuum energy in the presence of a background field,
and/or when the quantized field occupies a bounded spatial region
and is, therefore, subject to boundary conditions? This is,
precisely, the origin of the Casimir\cite{casimir} energy.

\bigskip

\section{Massless scalar field subject to Dirichlet boundary
conditions in one spatial direction}

Suppose we restrict our field to live between two parallel
plates, separated a distance $a$ in the $x$ direction, while
satisfying at the position of such plates
\[
\phi (t,0,y,z)=\phi(t,a,y,z)=0\,.
\]

In this case, the negative and positive frequency components of the
field are proportional to $\sin(k_n x)$, where $k_n=\frac{n\pi}{a}$,
with $n=1,2,...$ and\[\omega_n =(k_n^2+k_y^2+k_z^2)^{\frac12}\,.\]

So, the vacuum energy per unit area of the plates is given by
\[
\frac{E_V}{A}=\frac12 \int_{-\infty}^{\infty} \frac{dk_y
dk_z}{(2\pi)^2} \sum_{n=1}^{\infty} \left((\frac{n\pi}{a})^2
+k_y^2+k_z^2\right)^{\frac12}\,.
\]

As anticipated, this vacuum energy per unit transversal area is
divergent: both the series and the integral are so. In order to
give it an interpretation one must, as usual, regularize this
expression, isolate the divergencies and then renormalize
(whenever possible) the classical energy through physical
considerations.

The first regularization method we will employ is the one known as
zeta function regularization\cite{dowkerz,hawking} (it is based on
the analyticity properties of the zeta function of an operator,
which in this case is minus the Laplacian. A formal definition of the
spectral function known as zeta function will be given in the next
chapter). In this framework, we define
\begin{equation}
\frac{E_V}{A}=\frac{\mu}{2} \int_{-\infty}^{\infty}
\frac{dk_ydk_z}{(2\pi)^2} \sum_{n=1}^{\infty} ((\frac{n\pi}{a\mu})^2
+(\frac{k_y}{\mu})^2+(\frac{k_z}{\mu})^2)^{-\frac{s}{2}}\rfloor_{s=-1}\,.
\label{ez}
\end{equation}

Here, $s$ is a complex variable, with $\Re (s)$ big enough to
guarantee convergence. The result will, in this region, define an
analytic function of $s$. The vacuum energy will then be defined
through its analytic extension to $s=-1$. The parameter $\mu$, with
mass dimension, was introduced to make the quantity under the sum
dimensionless, and should disappear from any physically sensible
result.

Now, using\cite{abram}
\begin{equation}
z^{-s}=\frac{1}{\Gamma(s)} \int_{0}^{\infty} dt \,t^{s-1}e^{-zt},
\,\,\Re(s)>0\,\,,\,\Re(z)>0\,, \label{gam}
\end{equation}
equation (\ref{ez}) can be rewritten as
\[
\frac{E_V}{A}=\frac{\mu}{2} \int_{-\infty}^{\infty} \frac{dk_y
dk_z}{(2\pi)^2}
\sum_{n=1}^{\infty}\frac{1}{\Gamma(\frac{s}{2})}\int_{0}^{\infty}
dt\, t^{\frac{s}{2}-1}e^{((\frac{n\pi}{a\mu})^2
+(\frac{k_y}{\mu})^2+(\frac{k_z}{\mu})^2))t}\rfloor_{s=-1}\,.
\]

For $\Re(s)$ big enough, the sum and the integral can be reversed
to get
\[
\frac{E_V}{A}=\frac{\mu}{2}\sum_{n=1}^{\infty}\frac{1}{\Gamma(\frac{s}{2})}\int_{0}^{\infty}
dt \,t^{\frac{s}{2}-1}e^{-(\frac{n\pi}{a\mu})^2
t}\int_{-\infty}^{\infty} \frac{dk_y
dk_z}{(2\pi)^2}e^{-[(\frac{k_y}{\mu})^2+(\frac{k_z}{\mu})^2]t}\rfloor_{s=-1}\,.
\]

Now, both Gaussian integrals can be performed and equation
(\ref{gam}) can be used again, to obtain
\[
\frac{E_V}{A}=\frac{(\mu)^3}{8\pi}\frac{\Gamma(\frac{s}{2}-1)}{\Gamma(\frac{s}{2})}
\sum_{n=1}^{\infty}(\frac{n\pi}{a\mu})^{-s+2} \rfloor_{s=-1}=\]
\[
\frac{(\mu)^3}{4\pi (s-2)}(\frac{a\mu}{\pi})^{s-2}{\zeta}_{R} (s-2)
\rfloor_{s=-1} \,.
\]

Here, we have used the definition of Riemann's zeta function
\[
\zeta_R (s)=\sum_{n=1}^{\infty}n^{-s}\,,\,\Re(s)>1\,.
\]

The analyticity properties of Riemann's zeta are well known: The
previous series defines an analytic function for $\Re(s)>1$, and its
analytic extension to the whole $s$-plane presents only a simple
pole at $s=1$. In particular, its value at $s=-3$ is $\frac{1}{120}$.

So, our final result for the vacuum energy is
\begin{equation}
\frac{E_V}{A}=-\frac{\pi^2}{1440 a^3}\,. \label{e1}
\end{equation}

To summarize, the zeta function regularization gives, in this
simple case, a finite result, and no further renormalization is
needed.

\bigskip

Now, let us compare this result to the one given by another
regularization method: the exponential cutoff one (as we will see
later, it is based on the use of another spectral function, known
as the trace of the heat kernel). In this case, we define
\[
\frac{E_V}{A}=\frac{1}{2} \int_{-\infty}^{\infty}
\frac{dk_ydk_z}{(2\pi)^2} \sum_{n=1}^{\infty}
((\frac{n\pi}{a\mu})^2
+(\frac{k_y}{\mu})^2+(\frac{k_z}{\mu})^2)^{\frac12}e^{-((\frac{n\pi}{a\mu})^2
+(\frac{k_y}{\mu})^2+(\frac{k_z}{\mu})^2)^{\frac12}t}\rfloor_{t=0}\,.\]

Here, it is the exponential to insure convergence, thus allowing
for the interchange of sums and integrals. Again, the parameter
$\mu$, with units of mass, is arbitrary. The previous equation can
also be written as
\[
\frac{E_V}{A}=-\frac{\mu}{2}
\frac{d}{dt}\rfloor_{t=0}\int_{-\infty}^{\infty} \frac{dk_y
dk_z}{(2\pi)^2} \sum_{n=1}^{\infty} e^{-((\frac{n\pi}{a\mu})^2
+(\frac{k_y}{\mu})^2+(\frac{k_z}{\mu})^2)^{\frac12}t}=
\]
\[
-\frac{\mu}{4\pi} \frac{d}{dt}\rfloor_{t=0}\int_{0}^{\infty} dk k
\sum_{n=1}^{\infty} e^{-((\frac{n\pi}{a\mu})^2
+(\frac{k}{\mu})^2)^{\frac12}t}\,,
\]
or, after interchanging sum and integral and changing variables,
\[
\frac{E_V}{A}=-\frac{{\mu}^{3}}{2}
\frac{d}{dt}\rfloor_{t=0}\sum_{n=1}^{\infty}\int_{(\frac{n\pi}{a\mu})^2}^{\infty}
dk e^{-k^{\frac12}t}\,.\]

Now the sum can be evaluated through the Euler-Mc Laurin formula \be
\sum_{n=1}^{\infty} f(n) = -\frac12 f(0) +\int_{0}^{\infty} f(x)dx -
\sum_{k=1}^{\infty} \frac{1}{(2k)!} B_{2k} f^{(2k-1)}(0)\,.
\label{emcl} \ee

I will leave it as an exercise for you to show that the final
result for the vacuum energy in this regularization scheme is
given by
\begin{equation}
\frac{E_V}{A}=\frac{3a\mu^4}{2\pi^2
t^4}\rfloor_{t=0}-\frac{\mu^3}{4\pi
t^3}\rfloor_{t=0}-\frac{\pi^2}{1440a^3}\,. \label{e2}
\end{equation}

Notice that two divergencies remain, in the form of poles. The first
one is nothing but the vacuum energy in the whole space (it comes
from the integral in equation (\ref{emcl})). The second is due to the
mode n=0 (first term in the r.h.s. of the same equation) and, thus,
has its origin in boundary conditions. They can be elliminated
through the prescription $\frac{E_V}{A}\rightarrow0$, when
$a\rightarrow\infty$. This can, in fact, be understood as a
renormalization of the classical energy which is, in this problem,
of pure geometrical origin, and has the form
\[
E_{class} =paA+\sigma A\,,
\]
where $p$ is a pressure, and $\sigma$, a surface tension. The
remaining finite part is then in agreement with the result from
$\zeta$ regularization (equation (\ref{e1})). But our example is
simple (there is no mass, and the boundaries are flat ones). In the
rest of these lectures, we will study the connection between both
regularization methods and discuss the renormalization of Casimir
energies in more general cases.

\bigskip

{\it Exercise 1} - Complete the calculation of the vacuum energy in
the exponential cutoff regularization.

\bigskip

{\it Exercise 2} - Obtain Casimir's result for an electromagnetic
field between conducting plates. Hint: Only transversal modes
contribute after Gupta-Bleuler quatization. Solve Maxwell's
equations in the frame $k_y =0$. Show that, for $E_x =E_z =0$, one
scalar Dirichlet mode contributes while, for $B_x =B_z =0$, appart
from the Dirichlet mode, a constant one remains. So, going back to
arbitrary $k$
\[
E_V =\frac12 \int \frac{d^3 k}{(2\pi)^3}|k_{\parallel}| +
\sum_{n=1}^{\infty}\int \frac{d^3
k}{(2\pi)^3}\sqrt{(k_{\parallel}^2 + (\frac{n\pi}{a})^2)}\,,
\]
where $k_{\parallel}$ is the momentum parallel to the plates.

\chapter{Elliptic differential operators and boundary problems.
Spectral functions}

In the previous section, we have evaluated zero point energies
through two different regularization methods: the $\zeta$ function
and the exponential cutoff. They are based on the use of certain
functions of the spectrum of a given differential operator, called
spectral functions. In this section, I would like to discuss under
which conditions such functions can be defined, as well as some of
their useful properties.\footnote{Suggested bibliography:
\cite{seeley-sb,seeley-cb,seeley-trazas,agmon,libro-gilkey}}

\section{Differential operators on compact boundaryless manifolds}

Let $M$ be a compact boundaryless manifold of dimension $n$, and
$E$ a complex vector bundle over $M$.

\bigskip

{\bf{\cal Definition: Partial differential operator}}

A partial differential operator of order $m$ acting on sections of
$E$ can be written, in local coordinates, as
\[A=\sum_{|\alpha|\leq m} a_{\alpha}(x) D_x^{\alpha}\,,\] \[ {\rm
where}\, D_x^{\alpha}=\prod_{j=1}^n
\left(-i\frac{\partial}{\partial x_j}\right)^{\alpha _j} \quad
{\rm and}\, |\alpha|=\sum_{j=1}^n \alpha_j \,.\]

Here, the coefficients $a_{\alpha}(x)$ are, in general, $q\times q$
matrices.

\bigskip

{\it Example}: Consider the operator
$-\frac{d^2}{dx^2}+x\frac{d}{dx}+1$. In this case $n=1,m=2$. So
$j=1, |\alpha|=\alpha_1\leq2$. The operator can be written in
compact form as
\[
\sum_{\alpha_1\leq2} a_{\alpha_1}(x)
(-i)^{\alpha_1}(\frac{d}{dx_1})^{\alpha_1}\,.
\]

To $\alpha_1=0$ corresponds a coefficient $a_0(x)=1$; the coefficient
corresponding to $\alpha_1=1$ is $a_1(x)=ix$ and $a_2(x)=1$
corresponds to $\alpha_1=2$.

\bigskip

{\bf{\cal Definition: Symbol of a differential operator}}

The symbol of the operator $A$ is defined as
\[\sigma(A)=\sigma(x,\xi)=\sum_{|\alpha|\leq m} a_{\alpha}(x)
{\xi}^{\alpha}\,.\]

It is the polinomial of order $m$ in the dual variable $\xi$
obtained by formally replacing $D_x^{\alpha}$ by the monomial
${\xi}^{\alpha}$.

In terms of the symbol, the action of the operator on functions
in its domain can be written
\[
Af(x)=\int d\xi e^{ix.\xi}\sigma(x,\xi)\tilde{f}(\xi)\,.
\]
where $\tilde{f}(\xi)$ is the Fourier transform of $f(x)$.

{\it Example}: For the operator of the previous example
$\sigma(x,\xi)=\xi^2+ix\xi+1$. It is an easy exercise to show that
\[
(-\frac{d^2}{dx^2}+x\frac{d}{dx}+1)f(x)=\int d\xi
e^{ix.\xi}(\xi^2+ix\xi+1)\tilde{f}(\xi)\,.
\]

\bigskip

{\it Exercise 3} - Show that symbols compose according to the rule
\[
\sigma(PQ)=\sum_{\gamma} \partial_{\xi}^{\gamma}
(p(x,\xi))D_x^{\gamma}(q(x,\xi))/\gamma!\,,
\]
where $\gamma!=\gamma_1!\gamma_2!...\gamma_m!$ and $p(x,\xi),
q(x,\xi)$ are the symbols of $P$ and $Q$ respectively. Verify the
formula in the case $P=p(x)\frac{d}{dx}$ and $Q(x)=q(x)\frac{d}{dx}$.
Hint: start from Leibniz rule for the composition of operators.

\bigskip

{\bf{\cal Definition: Principal symbol}}

The principal symbol is the highest order part of the symbol.
It's a homogeneous polinomial of degree $m$ in $\xi$
\[\sigma _m (A)=\sigma _m (x,\xi)=\sum_{|\alpha|= m} a_{\alpha}(x)
{\xi}^{\alpha}\,.\]

In our example $\sigma _2 (x,\xi)=\xi^2$.

\bigskip

{\bf {\cal Definition: Ellipticity}}

The differential operator is said to be elliptic if its principal
symbol is invertible for $|\xi|=1$ (it has no zero eigenvalue for
$|\xi|=1$ or, equivalently, $det\, \sigma _m (x,\xi)\neq 0$ for
$|\xi|=1$).

This is obvious in our example.

\bigskip

{\bf {\cal Definition: Ray of minimal growth of the resolvent}}

Given an operator $A$, its resolvent is the operator
$(A-\lambda)^{-1}$.

The ray ${\cal K}=\{arg(\lambda)=\theta \}$ in the complex plane
$\lambda$ is called a ray of minimal growth of the resolvent if
there are no eigenvalues of the principal symbol on such ray, i.e.,
\[\sigma _m (x,\xi) u=\lambda u\]
has only the trivial solution for $\lambda \,\epsilon\,{\cal K}$.

\bigskip

It can be proved that, along such ray, the $L^2$ norm of the
resolvent is $O(\frac{1}{|\lambda|})$.

\bigskip

In our example, the problem $\xi^2 u=\lambda u$ has only the
trivial solution for any $\lambda\neq\xi^2$. Since $\xi$ is real,
any ray in $C-R^+$ is a ray of minimal growth.

\bigskip

\section{Complex powers of a differential operator}

\bigskip

Given an elliptic differential operator $A$, with a ray of minimal
growth ${\cal K}$ one defines, for $\Re(s)>0$
\[A^{-s}=\frac{i}{2\pi}\int_{\Gamma} {\lambda}^{-s}
(A-\lambda)^{-1} d\lambda\,,\] where $\Gamma$ is a curve starting at
$\infty$, coming along the ray ${\cal K}$ to a small circle at the
origin, and back to $\infty$ along the ray. (Notice such curve
encloses the eigenvalues of the principal symbol in a clockwise
sense).

To describe $A^{-s}$, one can construct an approximation
$B(\lambda)$, to the resolvent $(A-\lambda)^{-1}$, known as the
parametrix, wich reproduces the behaviour of the resolvent as
$\lambda\rightarrow\infty$ along the ray of minimal growth
\cite{seeley-sb}.

The parametrix is constucted by considering $\lambda$ as part of the
principal symbol of the operator $A$, proposing
\[
\sigma(B)\sim\sum_{j=0}^{\infty}b_{-m-j}(x,\xi,\lambda)\,\] and
imposing
\[
\sigma(B(A-\lambda))=I\,.\] The coefficients $b_{-m-j}$ are known
as Seeley's coefficients, and they can be seen (from the formula
for the composition of symbols) to satisfy the following set of
algebraic equations
\[
b_{-m}(a_m-\lambda)=I
\]
\[
b_{-m-l}(a_{m}-\lambda)+\sum({\partial}_{\xi}^{\alpha}b_{-m-j})(D_x^{\alpha}a_{m-k})=0\quad
\rm{for} \,l>0\,.
\]

Here, the sum must be taken over all $k+j+|\alpha|=l$ and $j<l$.
The $a_{m-k}$ are the diverse order symbols of the differential
operator $A$.

From this coefficients, an approximation to the symbol of $A^{-s}$ is
\[\sigma(A^{-s})\sim\sum_{j=0}^{\infty} \frac{i}{2\pi}\int_{\Gamma}
{\lambda}^{-s} b_{-m-j}(x,\xi ,\lambda ) d\lambda\,,\]

\bigskip

Starting from this expression, it can be shown that, for
$\Re(ms)>n$, the kernel $K_{-s}(x,y)$ of $A^{-s}$ is continuous. For
$x\neq y$, it extends to an entire function of $s$. For $x=y$, it
extends to a meromorphic function, whose only singularities are
simple poles at $s=\frac{n-j}{m}, \, j=0,1...$. Each pole is due to
a particular term in the previous expression, and the residues are
thus determined by the integrals along $\Gamma$ of Seeley's
coefficients.

\bigskip

\section{$\zeta$ function. Relation to eigenvalues}

{\bf {\cal Definition: $\zeta$ function}}

Given an elliptic operator $A$, the first spectral function we will
consider is its $\zeta$ function, defined as the trace of its
$(-s)$-th power
\[\zeta(A,s)=tr(A^{-s})\,.\]

The analyticity properties of the $\zeta$ function are derived from
those of $K_{-s}(x,x)$ (see last paragraph in the previous section).
The residues at the poles of $\zeta(A,s)$ are the integrals, over the
manifold $M$, of the residues corresponding to $K_{-s}(x,x)$, and
are thus determined by Seeley's coefficients.

\bigskip

When the operator has a complete orthogonal set of eigenfunctions,
its $\zeta$ function can be expressed in terms of the corresponding
eigenvalues.

In fact, suppose the bundle has a smooth Hermitian inner product, and
$M$ a smooth volume element $dv$. If the operator $A$ is normal with
respect to these structures ($A^{\dag}A=AA^{\dag}$), then it has a
complete orthonormal set of eigenfunctions
$A\,{\phi}_k={\lambda}_k\,{\phi}_k$, and one can write
\[K_{-s}(x,y)\,dv_y=\sum {\lambda}_k^{-s} {\phi}_k (x) {\phi}_k^{\dag}
(y)\, dv_y\,.\]

Now, taking $x=y$ and integrating over $M$,  \be
\zeta(A,s)=tr(A^{-s})=\sum {{\lambda}_k}^{-s}\,.\label{uno}\ee

\bigskip

{\it Example}: Consider, on the unit circle, the operator
\[
A=-\frac{d^2}{dx^2}+P\,, \] where $P$ is the projector on zero
modes. Its eigenvalues are given by $\lambda_n=n^2$, for $n=\pm
1,\pm 2,...$, and $\lambda_0 =1$ for the zero mode. So
\[
\zeta(A,s)=2\sum_{n=1}^{\infty}(n^2)^{-s}+1=2\zeta_R(2s)+1 \,. \]

Now, this Riemann's $\zeta$ function is known to be analytic for
$\Re(2s)>1$, i.e., $\Re(s)>\frac12=\frac{n}{m}$. Its analytic
extension presents a simple pole at $2s=1$.

\bigskip

{\it Exercise 4} - Show that, in this case, the only nonvanishing
Seeley's coefficient is $b_{-2}=(\xi^2-\lambda)^{-1}$. This
coefficient gives rise to the unique pole at $s=1$. Determine the
residue by integrating $b_{-2}$.

\bigskip

\section{The heat kernel and its trace}

If the eigenvalues of the principal symbol all lie in a region $S_0$
:$-\frac{\pi}{2}+\varepsilon<arg(\lambda)<\frac{\pi}{2}-\varepsilon$,
then the spectrum of $A$ lies in the sector $S_{\alpha}$
:$-\frac{\pi}{2}+\varepsilon<arg(\lambda +\alpha
)<\frac{\pi}{2}-\varepsilon$ for some $\alpha>0$, and one can define
\[ e^{-At}=\frac{i}{2\pi}\int_{\Gamma} e^{-\lambda t}
(A-\lambda)^{-1} d\lambda \quad,\,t>0 \] where $\Gamma$ is the
border of $S_{\alpha}$. It can be shown that $e^{-A t}$ is the
fundamental solution of the heat equation $A u +\frac{\partial u
}{\partial t}=0$, with $u(x,0)=\delta(x)$. For this reason, it is
called the heat kernel of the operator $A$.

The approximation to the resolvent then allows for the limit
$t\rightarrow {0^+}$ in the previous integral, and one thus obtains
an asymptotic expansion for $e^{-A t}$ in increasing (in general,
noninteger) powers of t. The coefficients in such expansion are also
determined by Seeley's coefficients.

As before, if $A$ has a complete set of eigenfunctions, the kernel of
$e^{-At}$ can be written as
\[K(t,x,y)= \sum_k e^{-{\lambda}_k t} {\phi}_k (x) {\phi}_k^{\dag}
(y)\,.\]

and its trace
\be h(A,t)=tr ( e^{-At})=\sum_k e^{-{\lambda}_k
t}\,.\label{dos}\ee

This is the definition of the trace of the heat kernel, the second
spectral function we will be using.

\bigskip

There is a very close relationship between the $\zeta$ function of
an operator and the trace of its heat kernel. In fact, from
equations (\ref{uno}) y (\ref{dos})
\be \zeta(A,s)=\sum_k
{{\lambda}_k}^{-s}=\frac{1}{\Gamma(s)} \int_0^{\infty} dz\,\sum_k
e^{-{\lambda}_k z}z^{s-1}=\frac{1}{\Gamma(s)} \int_0^{\infty}
dz\,h(A,z)z^{s-1}\,.\ee

Both spectral functions are related through a so called Mellin
transform.

\bigskip

{\it Exercise 5} - Obtain the trace of the heat kernel for the
operator in the previous exercise. Show that its $\zeta$ function
is, in fact, the Mellin transform of the trace of the heat kernel.

\section{Elliptic boundary systems}

Up to this point, we have considered boundaryless manifolds. How do
the concepts we have introduced extend to manifolds with boundaries?
Let $M$ be a compact manifold of dimension $n$, with a smooth
boundary $\partial M$.

In each local coordinate system, call $x=(x_1,...,x_{n-1})$ the
coordinates on $\partial M$. Let $t$ ($\epsilon {\bf R}$) the
interior normal to the boundary. So, $(x,t)\epsilon{\bf R}^n$. Call
${\bf R}_+ ^n$ the half space $t\geq 0$. We will consider, in ${\bf
R}_+ ^n$, the differential operator of order $m$:  \[A=\sum _{j=0}^m
A_j (x,t) D_{t}^{m-j}\,,\quad(D_t=-i\frac{\partial}{\partial t})\,,\]
where $A_j$ is a differential operator of order $\leq j$ on ${\bf
R}^{n-1}$.

Then, calling $(\xi,\tau )$ the symbolic variable corresponding to
$(x,t)$, we have:
\[\sigma (A)=\sum _j \sigma (A_j)(x,t,\xi ) {\tau}^{m-j}\,.\]

The principal symbol is
\[{\sigma}_m=\sum _j {\sigma}_j (A_j)(x,t,\xi ) {\tau}^{m-j}\,.\]

Moreover, we define a partial principal symbol, by:
\[{{\sigma}_m}^{\prime}= \sum _j {\sigma}_j (A_j)(x,0,\xi )
{D_t}^{m-j}\,.\]

Suppose, near the boundary, we have certain given operators
(defining boundary conditions)
\[B_j=\sum_{k=1}^m B_{jk} D_t ^{m-k}\,,\quad 1\leq j\leq \frac{m q}{2}\] where the $B_{jk}$
are a system of differential operators ($1\times q$ matrices) acting
on ${\bf R}^{n-1}$. We will concentrate on the case in which these
boundary operators are merely multiplicative. Then,
\[\sigma (B_j)=\sum_{k=1}^m \sigma
(B_{jk}) {\tau} ^{m-k}\,,\] \[{\sigma}^{\prime} (B_j)=\sum_{k=1}^m
{\sigma}(B_{jk}) {D_t} ^{m-k}\,.\]

{\bf {\cal Definition: Elliptic boundary system}}

The collection of operators $A, B_1,..., B_{\frac{mq}{2}}$
constitute an elliptic boundary system if $A$ is elliptic and, for
$g=(g_1,..., g_{\frac{mq}{2}})$ arbitrary, $x$ in ${\bf R}^{n-1}$ and
$\xi \neq 0$ in ${\bf R}^{n-1}$ there is a unique solution to the
following problem on $\{t>0\}$ :
\[{{\sigma}_m}^{\prime}(A)(x,\xi,D_t) u=0\] \[\lim _{t\rightarrow \infty}
u(t)=0\] \[{\sigma}^{\prime} (B_j)(x,\xi,D_t) u=g_j \quad {\rm at} \,
t=0\quad {\rm for}\,j=1,...,\frac{mq}{2}\,.\] This condition is also
known as the Lopatinski-Shapiro condition \cite{libro-gilkey}. When
it holds, an operator $A_B$ can be defined as the operator $A$,
acting on functions $u$ such that $B_ju=0$.

\bigskip

{\bf {\cal Definition: Strong ellipticity}}

The collection $A, B_1,..., B_{\frac{mq}{2}}$ constitutes a strongly
elliptic boundary system in a cone ${\cal K}\,\subset\,{\bf C}$
including the origin if

$i$) For $(\xi,\tau)\neq (0,0)$ ${\sigma}_m (A)$ has no eigenvalue
in ${\cal K}$ and

$ii$) For each $x$ and each $(\xi , \lambda )\neq (0,0)$, with
$\lambda\,\epsilon\,{\cal K}$, the boundary problem
\[{{\sigma}_m}^{\prime}(A)(x,\xi,D_t) u=\lambda u\] \[\lim
_{t\rightarrow \infty} u(t)=0\] \[{\sigma}^{\prime} (B_j)(x,\xi,D_t)
u=g_j \quad {\rm at} \, t=0\,,j=1,...,\frac{mq}{2}\] has a unique
solution. (Note this reduces to the Lopatinski-Shapiro condition for
$\lambda=0$). The cone ${\cal K}$ is known as Agmons cone
\cite{agmon}.

When the strong ellipticity condition holds, an approximation to the
resolvent $(A_B -\lambda )^{-1}$ can be found \cite{seeley-cb} and,
from it, one can obtain
\[(A_B)^{-s}=\frac{i}{2\pi}\int_{\Gamma} {\lambda}^{-s}
(A_B-\lambda)^{-1} d\lambda\,,\] with $\Gamma$ an appropriate curve
in the cone where $(A_B-\lambda)^{-1}$ is known to exist. The
coefficients in the expansion of the parametrix must now satisfy not
only the condition of representing an inverse for $A_B-\lambda$, but
must also adjust the boundary condition. Then, appart from the
volume Seeleys coefficients $b_{-m-j}$, there are new boundary
coefficients $d_{m-j}$, and their determination leads to a set of
differential (rather than algebraic) equations.

The conclusions concerning the pole structure of $K_{-s}(x,y)$ are
similar to those in the boundaryless case. However, in this case,
the residues at the poles are given by volume integrals of the
coefficients $b$, plus boundary integrals of the new ones, $d$. The
zeta function can, as before, be defined as the trace of the
$(-s)$-th power. In particular, if the operator $A_B$ has a complete
set of eigenfunctions

\be \zeta(A,s)=\sum_k ({\lambda}_k)^{-s}\,,\ee \be h(A_B,t)=\sum_k
e^{-{\lambda}_k t}\,,\ee
and both spectral functions are again
related through a Mellin transform.

\bigskip

{\it Example}: The Laplacian on the boundary of a cylinder with
Dirichlet boundary conditions

Let $A=-\partial_x^2-\partial_y^2$, acting on functions
$\varphi(x,y)$ such that $\varphi(0,y)=\varphi(1,y)=0$, and periodic
in the $y$-direction. Then, the boundary corresponds to $x=0,1$. The
boundary operator is merely multiplicative, $B=1$. Let $(\tau,\xi)$
be the Fourier variables associated with $(x,y)$. Then
\[
\sigma(A)=\sigma_2 (A)=\xi^2+\tau^2
\]
and \[\sigma(B)=\sigma_0 (B)=\sigma_0^{\prime}(B)=1
\]

Then, one can show that

1) The differential operator is elliptic

In fact, $\sigma_2 (A)=\xi^2+\tau^2\neq 0$, for $(\tau,\xi)\neq
(0,0)$, since $\tau, \xi \epsilon\,{\bf R}$.

2) The boundary problem is elliptic in the weak (Lopatinski-Shapiro)
sense

Let us consider the boundary at $x=0$. Then $x$ is the variable
normal to the boundary (t). The differential equation \[
\sigma_2^{\prime} (A)\,u=(-\partial_x^2+\xi^2)\,u=0\] has, for
$\xi\neq 0$, solutions of the form
\[
u=C\exp(|\xi|x)+D\exp(-|\xi|x)\,.
\]

The condition that these solutions vanish for $x\rightarrow \infty$
requires $C=0$. For the remaining part, the problem
\[
u(0,\xi)=g\] has, for arbitrary $g$, the unique solution $D=g$,
which shows that weak ellipticity holds at $x=0$. It can similarly
be shown to hold at $x=1$.

3) The differential operator has an Agmon cone

The equation
\[
\sigma_2 \,u=(\tau^2+\xi^2)\,u=\lambda\,u \,,\] with $(\tau,\xi)\neq
(0,0)$ has nontrivial solutions only for
$\lambda=\tau^2+\xi^2\,\epsilon\,{\bf R}^{+}$. So, the differential
operator has an Agmon cone given by ${\cal K}={\bf C}-{\bf R}^{+}$.

4) The boundary problem is strongly elliptic (has an Agmon cone)

Again, we consider the boundary at $x=0$. The differential equation

\[
\sigma_2^{\prime} (A)\,u=(-\partial_x^2+\xi^2)\,u=\lambda\,u\] has,
for $(\xi,\lambda)\neq (0,0)$, solutions of the form
\[
u=C\exp(\sqrt(\xi^2-\lambda^2)x)+D\exp(-\sqrt(\xi^2-\lambda^2)x)\,.
\]

The condition that they vanish for $x\rightarrow \infty$ requires
$C=0$. For the remaining part, the problem
\[
u(0,\xi,\lambda)=g\] has, for arbitrary $g$, the unique solution
$D=g$, which, together with 3), shows that the boundary problem is
strongly elliptic in ${\cal K}={\bf C}-{\bf R}^{+}$.

\chapter{Comparison of zeta and exponential regularizations of Casimir
energies}

In this chapter we show that, in general, both the $\zeta$ and the
exponential regularization give divergent results for the Casimir
energy. A general relationship between both results is established,
and the existence of a unique, physically meaningful result after
renormalization is discussed.\footnote{Suggested bibliography:
\cite{nos1,grubb}}

\section{General result}

We have, in the first section, studied a very simple example of
Casimir energy evaluation: a massless scalar field between
``conducting'' plates. This is simple in two senses: the field is
massless and the geometry is a planar one. In this case, we obtained
a finite result for the vacuum energy in the $\zeta$ regularization,
and divergencies in the form of poles in the exponential cutoff one.
Moreover, these divergencies showed a dependence with the distance
between plates consistent with the classical action, and different
from that of the finite part. So, they could be renormalized away.

Now, we face some questions: Is renormalization still possible in a
more general case? Does the $\zeta$ function always give a finite
result? Are the divergencies in the exponential regularization
always poles? What is, in a more general case, the relationship
between the results given by both regularizations?

To answer these questions, we will use the general results in chapter
2, concerning the structure of $\zeta$ and trace of the heat kernel,
as applied to second order operators.

Recall that, for instance, in the scalar case, the vacuum energy is
given by
\[ E_V=\frac 12\sum_n\omega _n\,,
\label{E}
\]
where $\omega _n$ are the zero point energies. Consider a scalar
field, living in a $d+1$-dimensional space-time, where $d$ is the
dimension of the compact spatial manifold, with or without a smooth
boundary.

Then,
\[ \omega _n=\lambda _n^{1/2}\qquad, \] where the $\lambda _n$
satisfy the associated boundary problem
\begin{equation} \label{op}D_B\varphi _n=\left\{ {D\varphi _n=\lambda
_n\varphi _n}\atop{%
B\varphi _n=0} \right. \qquad.  \end{equation}

$D$ is a second order differential operator on $M$;  $B$ is the
operator defining boundary conditions. In what follows, we will
refer to the boundary value problem (\ref{op}) as $(D,B)$.

The $\zeta$-regularized \cite{dowkerz,hawking} vacuum energy is
defined as
\begin{equation} \label{z} \begin{array}{c} E_\zeta \equiv \left. \frac
\mu 2\sum\limits_n\left( \frac{\lambda _n}{\mu ^2}\right) ^{-\frac
s2}\right\rfloor _{s=-1}=\left.  \frac \mu 2\zeta \left(\frac s2,
\frac{D_B}{\mu ^2}\right)  \right\rfloor _{s=-1} \\ \\ =\left. \frac
\mu 2Tr\left( \frac{D_B}{\mu ^2}\right) ^{-\frac s2}\right\rfloor
_{s=-1}
\end{array}\,\label{erz} \end{equation}
while the cutoff regularized expression is given by
\begin{equation} E_{\exp }\equiv \left. \frac \mu
2\sum\limits_n\frac{\lambda _n^{1/2}}\mu e^{-t\frac{\lambda
_n^{1/2}}\mu }\right\rfloor _{t=0}=\left. -\frac \mu 2\frac
d{dt}\left( h\left( t,\frac{{D_B}^{\frac12}}{\mu}\right) \right)
\right\rfloor _{t=0}\qquad, \label{erc}
\end{equation} where
\begin{equation} \label{h} h\left( t,\frac{{D_B}^{\frac12}}{\mu}\right)
=\sum\limits_ne^{-t\frac{ \lambda _n^{1/2}}\mu }=Tr\left(
e^{-\frac t\mu D_B^{1/2}}\right)  \qquad.  \end{equation}

Remember $\mu $ is, in both cases, an arbitrary parameter with
dimensions of mass.

In order to study the connection between both regularizations, we
will make use of the following well known result

\bigskip\

{\bf {\cal Lemma 1}}\cite{seeley-trazas}

Let $D$ be a second order differential operator, acting on a smooth
compact $d$-dimensional manifold $M$, and let $B$ be the differential
operator defining boundary conditions on $\partial M$. If the
boundary problem (D,B) is strongly elliptic with respect to
$C-R^{+}$, then

a)  \[ \mu^{-s}\Gamma \left( \frac s2\right) \zeta \left( \frac
s2,\frac{D_B}{\mu ^2} \right) \] is analytic for $\Re(s)>d$, and it
extends to a meromorphic function, with the following singularity
structure
\[ \label{z1} \mu^{-s}\Gamma \left( \frac s2\right) \zeta \left(
\frac s2,\frac{D_B}{\mu ^2}\right) =\sum\limits_{j=0}^N
\frac{2a_j}{s+j-d}+r_{N}\left( \frac s2\right) \qquad,
\] where $r_{N}\left( \frac s2\right) $ is analytic for $\Re(s)>d-N-1$.

b) For each real $c_1,c_2$ and each $\delta <\theta _0$,
\[ \label{propz}\left| \mu^{-s}\Gamma \left( \frac
s2\right) \zeta \left( \frac s2, \frac{D_B}{\mu ^2}\right) \right|
\leq C\left( c_1,c_2,\delta \right) e^{-\delta \left| \Im\left(\frac
s2\right)\right| }\quad ,\left|\, \Im\left(\frac s2\right)\right|
\geq 1,\,c_1\leq \Re\left(\frac s2\right)\leq c_2 \,.\] \bigskip

The coefficients $a_j$ in a) are determined by the integrated
Seeleys coefficients.

This Lemma clearly shows that the vacuum energy evaluated through
$\zeta$ regularization (equation (\ref{erz})) presents a singularity,
in the form of a pole, as long as $a_{d+1}\neq 0$. What can be said
about the exponential cutoff regularized expression (\ref{erc})? To
study its behaviour, we will prove the following

{\bf {\cal Lemma 2}}

Under the hypothesis of the previous Lemma

\bigskip

a)$\frac{d\,h\left( t,\frac{{D_B}^{\frac12}}{\mu}\right)}{dt}
=\frac{d\,Tr\left( e^{-\frac {t}{\mu} D_B^{1/2}}\right)}{dt}
=\sum\limits_n{-\frac{\lambda _n^{1/2}}{\mu }}\,e^{-t\frac{\lambda
_n^{1/2}}{\mu }}$ has an asymptotic expansion

\[\frac{d\,h\left( t,\frac{{D_B}^{\frac12}}{\mu}\right)}{dt}= \sum\limits_{k=0}^d
(-k)\frac{1}{2\mu}\frac{\Gamma \left( \frac{k+1}2\right) }{\Gamma \left(
\frac 12\right)} a_{d-k}\left( \frac{t}{2\mu}\right) ^{-k-1}+\] \[
\sum\limits_{k=1}^K (-k)\frac{1}{2\mu} \frac{\Gamma \left( -k+\frac
12\right) }{\Gamma \left( \frac 12\right) } 2a_{d+2k}\left(
\frac{t}{2\mu}\right)  ^{2k-1} + \] \[\sum\limits_{k=0}^K (2k+1)
\frac{1}{2\mu} \frac{\left( -1\right) ^{k}}{\Gamma \left( \frac 12\right)
\Gamma \left( k+1\right) }\left( \frac{t}{2\mu}\right) ^{2k} \left[
r_{d+2k+1}(-k-\frac 12) +\right. \] \[ a_{d+2k+1}\left( \Psi (1)
+\sum\limits_{l=0}^{k-1}\frac 1{k-l}\right) +\] \be \left.
\sum\limits_{j=0}^{d+2k}\frac{2a_j}{j-d-2k-1}+2a_{d+2k+1} \left((2k+1)\ln(
\frac{t}{2\mu})- \frac{1}{2k+1}\right)  \right]+\,\rho_K(t)\label{expdh}
\end{equation} where $\rho_K$ is
$O\left((\frac{t}{2\mu})^{2K+1+\varepsilon}\right)\,,
0<\varepsilon<1$ for $t\rightarrow 0$.

The important point here is that this asymptotic expansion contains,
for $t\rightarrow0$, not only poles (coming from the first sum), but
also logarithmic divergencies (coming from $k=0$ in the last one).
In fact, it presents poles of order $d+1,d,...,-1$, with
coefficients determined by $a_{d-k}, k=0,...,d)$. As to the
coefficient of the logarithm, it is determined by $a_{d+1}$.

\bigskip\

{\bf {\cal Proof}}

I will only sketch the proof here. For details, see $\cite{nos1}$.

Note, in the first place, that
\[ \Gamma \left(
s\right) \zeta \left( \frac s2,\frac{D_B}{\mu ^2}\right)
=\int_0^\infty t^{s-1}h\left( t,\frac{{D_B}^{\frac12}}{\mu}\right) dt
\] is the Mellin transform of $h\left(
t,\frac{{D_B}^{\frac12}}{\mu}\right) $.

It can also be written as
\[ \Gamma \left( s\right) \zeta \left( \frac s2, \frac{D_B}{\mu ^2}\right)
=\frac{\Gamma \left( s\right) }{\Gamma \left( \frac s2\right)
}\left[ \Gamma \left( \frac s2\right) \zeta \left( \frac s2,
\frac{D_B}{\mu ^2}\right) \right]=\] \be \frac{2^{s-1}}{\sqrt{\pi
}}\Gamma \left( \frac{s+1}2\right)  \left[ \Gamma \left( \frac
s2\right) \zeta \left( \frac s2,\frac{D_B}{\mu ^2}\right) \right]\,.
\label{z22} \ee

From {\cal Lema 1 }a), and the well known singularity structure of
$\Gamma \left( \frac{s+1}2\right) $, it turns out that (\ref{z22}) is
analytic for $\Re (s)>d$, and
\begin{equation} \label{h1}h\left( t,\frac{D_B}{\mu ^2}\right)
=\frac 1{2\pi i}\int\limits_{c-i\infty }^{c+i\infty }ds
\left(\frac{t}{\mu}\right)^{-s}\frac{2^{s-1}}{\sqrt{\pi }}\Gamma
\left( \frac{ s+1}2\right) \left[ \mu^{-s}\Gamma \left( \frac
s2\right) \zeta \left( \frac s2, \frac{D_B}{\mu ^2}\right) \right]
\,, \end{equation} where the integration path is such that $c>d$.

This expression can be derived, to obtain $\frac{dh}{dt}$
\[\frac{d\,h\left( t,\frac{D_B}{\mu ^2}\right)}{dt} =\] \[
\label{dh1} \frac 1{2\pi i}\int\limits_{c-i\infty }^{c+i\infty }ds
\frac{(-s)}{\mu}\left(\frac{t}{\mu}\right)^{-s-1}\frac{2^{s-1}}{\sqrt{\pi
}}\Gamma \left( \frac{ s+1}2\right) \left[ \mu^{-s}\Gamma \left(
\frac s2\right) \zeta \left( \frac s2, \frac{D_B}{\mu ^2}\right)
\right]\,,
\] where the integral is performed along the same curve as before.

Now, using {\cal Lemma 1} b), together with the fact that $\Gamma
\left( \frac{s+1}2\right) $ is $O\left( e^{\left( -\frac \pi
2+\epsilon \right) \left| Im\frac s2\right| }\right) $, for any
$\epsilon
>0$, it is possible to obtain an asymptotic expansion for
$\frac{d\,h\left( t,\frac{D_B}{\mu ^2}\right)}{dt}$ by moving the
integration path $\left( \ref{h1}\right) $ through the poles of
$\Gamma \left( \frac{s+1}2\right) \left[ \mu^{-s}\Gamma \left( \frac
s2\right) \zeta \left( \frac s2,\frac{D_B}{\mu ^2}\right) \right] $.
Such poles are located at $s=d-j$.

For $s=d-j=k\geq 0\quad \left( j\leq d\right) $ they are simple
poles, and contribute to the Cauchy integral with \[
\label{res1}\frac{-k}{2\mu}\frac{\Gamma \left( \frac{k+1}2\right)
}{\Gamma \left( \frac 12\right)}a_{d-k}\left( \frac{t}{2\mu}\right)
^{-k-1}\qquad ,\ k=0,1,\ldots ,d\,.  \]

For $s=d-j=-2k$ \quad $\left( k=1,2,\ldots \right) $ they are also
simple poles, and their contribution is \[
\label{res2}-k\frac{1}{2\mu}\frac{\Gamma \left( -k+\frac 12\right)
}{\Gamma \left( \frac 12\right)}2a_{d+2k}\left( \frac{t}{2\mu}\right)
^{2k-1}\qquad ,k=1,2,\ldots\,.  \]

For $s=d-j=-\left( 2k+1\right) \quad \left( k=0,1,\ldots \right)  $
they are simple and double poles, and give a contribution
\[ \label{res3}
\frac{(2k+1)}{2\mu}\frac{(-1)^k}{\Gamma \left( \frac 12\right)
\Gamma \left( k\right) }\left( \frac{t}{2\mu}\right) ^{2k}\left[
r_{d+2k+1}(-k-\frac 12)
+\sum\limits_{j=0}^{d+2k}\frac{2a_j}{j-d-2k-1}\right] \]
\[ \label{res4}\frac{(2k+1)}{2\mu}\frac{\left( -1\right)
^k}{\Gamma \left( \frac 12\right) \Gamma \left( k+1\right)}\left(
\frac{t}{2\mu}\right) ^{2k}a_{d+2k+1}\left[ 2\ln( \frac{t}{2\mu})-2
+\Psi(1) +\sum\limits_{l=0}^{k-1}\frac 1{k-l}\right] \] (The sum over
$l$ must be included whenever it makes sense).

So, moving the integration path in (\ref{h1}) till the singularity at
$s=-(2K+1)$ is included, we have

\[ \frac{d\,h\left( t,\frac{D_B}{\mu ^2}\right)}{dt}= \sum\limits_{k=0}^d
(-k)\frac{1}{2\mu}\frac{\Gamma \left( \frac{k+1}2\right) }{\Gamma \left(
\frac 12\right)} a_{d-k}\left( \frac{t}{2\mu}\right) ^{-k-1}+\] \[
\sum\limits_{k=1}^K (-k)\frac{1}{2\mu} \frac{\Gamma \left( -k+\frac
12\right) }{\Gamma \left( \frac 12\right) } 2a_{d+2k}\left(
\frac{t}{2\mu}\right)  ^{2k-1} +\] \[ \sum\limits_{k=0}^K (2k+1)
\frac{1}{2\mu} \frac{\left( -1\right) ^{k}}{\Gamma \left( \frac 12\right)
\Gamma \left( k+1\right) }\left( \frac{t}{2\mu}\right) ^{2k} \left[
r_{d+2k+1}(-k-\frac 12) +\right. \] \[a_{d+2k+1}\left( \Psi (1)
+\sum\limits_{l=0}^{k-1}\frac 1{k-l}\right) +
\sum\limits_{j=0}^{d+2k}\frac{2a_j}{j-d-2k-1}\,+ \] \be \left.
2a_{d+2k+1} \left((2k+1)\ln( \frac{t}{2\mu})- \frac{1}{2k+1}\right)
\right]+\,\rho_K(t)\,.  \end{equation}

The rest $\rho _K\left( t\right) $ is given by an integral as
(\ref{dh1}), but with $c<-2(K+1)$. As a result of {\cal Lema 1} b)
and the estimate for $\left| \Gamma \left( \frac{s+1}2\right)
\right|$ already discussed, this rest is $O\left( \left|
\frac{t}{2\mu}\right| ^{2K+1+\varepsilon }\right) $, which completes
the proof.
\bigskip

When evaluated at $t=0$, this asymptotic expansion gives, for the
exponentially regularized vacuum energy (\ref{erc})
\[ E_{\exp }=\left. -\frac{\mu}{2} \frac{dh\left( t,\frac{D_B}{\mu
^2}\right) }{dt}\right\rfloor _{t=0}=-\frac 1 2\left.
\sum\limits_{k=1}^d\left( -k\right) \frac{\Gamma \left( \frac{k+1}2\right)
}{ 2^{-k}\Gamma \left( \frac 12\right)  }a_{d-k}\left(\frac{t}{\mu}\right)
^{-k-1}\right\rfloor _{t=0} \, - \] \be \frac{1} {4\Gamma \left( \frac
12\right) }\left[ r_{d+1}\left( -\frac 12\right)  +a_{d+1}\left( \Psi
\left( 1\right) -2\right) +2\sum\limits_{j=0}^{d}\frac{%
a_j}{j-d-1}\right] +\frac1 2\frac{a_{d+1}}{\Gamma \left( \frac
12\right) } \ln \left( \frac{t}{2\mu}\right)\,. \label{Eexp}
\end{equation}

Going back to (\ref{z1}), the $\zeta$ regularized vacuum energy is
given by

\[ E_\zeta =\left. \frac \mu 2\zeta \left( \frac s2, \frac{D_B}{\mu
^2}\right) \right\rfloor _{s=-1} =\] \[\frac 1 {2\Gamma \left(
-\frac 12\right) }\sum\limits_{j=0}^{d}\frac{2a_j}{j-d-1}+\frac 1
{2\Gamma \left( -\frac 12\right) }r_{d+1}\left( -\frac 12\right)
\\ \\ +\left. \frac{\mu^{s+1}} {\Gamma \left( \frac s2\right) }
\frac{a_{d+1}}{s+1}\right\rfloor _{s=-1}=\] \[-\frac 1 {2\Gamma
\left( \frac 12\right) }\sum\limits_{j=0}^{d}
\frac{a_j}{j-d-1}-\frac 1 {4\Gamma \left( \frac 12\right)
}r_{d+1}\left( -\frac 12\right) +\] \be\frac 1 {2\Gamma \left(
\frac 12\right) }a_{d+1}\left( \frac{ \Psi \left( 1\right)
}2+1-\ln \left(2\mu\right)\right)  -\left. \frac 1 {2\Gamma \left(
\frac 12\right)  }\frac{a_{d+1}}{s+1} \right\rfloor _{s=-1}
\,.\label{Ez}
\end{equation}

From (\ref{Eexp}) and (\ref{Ez}) the following conclusions can be
stated:

Both regularization methods give, in principle, divergent results.

1) If $a_{d+1}$ vanishes, the $\zeta$ regularization gives a finite
result, which coincides with the minimal finite part in the
exponential regularization. This last method presents poles of orders
2, 3, $\ldots $, $d+1$. The residue at the pole of order $k+1$ is
given by the product of $\Gamma \left( k+1\right)$ by the residue of
$\frac {\mu}{2}\zeta \left( \frac {s}{2},\frac{D_B}{\mu ^2}\right)$
at $s=k\,\left( k=1,\ldots ,d\right)$.

2) In the general case, $\left( a_{d+1}\neq 0\right)$, the
exponential regularization shows, appart from the poles, a
logarithmic divergence, whose coefficient is minus the residue of
$\frac {\mu}{2}\zeta \left( \frac{s}{2},\frac{D_B}{\mu ^2}\right)$ at
$s=-1$. As a consequence, the difference between the minimal finite
part in the exponential regularization and the one in the $\zeta$
regularization is given by
\begin{equation} \label{dif} -\frac 1 2\frac{a_{d+1}}{\sqrt{\pi
}}\Psi \left( 1\right)  =\frac 1 2\frac{a_{d+1}}{\sqrt{\pi
}}\gamma\,, \end{equation} where $\gamma $ is the Euler-Mascheroni
constant. Both schemes show a logarithmic dependence on $\mu$ (as
discussed in \cite{blaue} for the $\zeta$ case). If the difference
between both results consists of renormalizable terms, a physical
interpretation will be possible, and all dependence on $\mu$ will
disappear.

All these results are also valid in the case of boundaryless
manifolds.

\bigskip

\section{Example: Massive scalar field in $1+1$}

You can now go back to our example of the massless scalar in $d+1=4$,
and test it falls into case 1).

As a simple example of case 2), consider a massive scalar field in
$d+1=2$ dimensions, satisfying periodic boundary conditions in the
spatial direction \[\varphi \left( t,L\right) =\varphi \left(
t,0\right)\,.\]

It is easy to see that
 \[\omega _{n}=\left[ m^2+\left( \frac{2n\pi
}{L}\right) ^2\right] ^{1/2}\, \,,n\epsilon Z. \]

Then, through $\zeta$ regularization,
\[
E_{\zeta}=\frac{\mu}{2}  \sum_{n=-\infty}^{\infty}
((\frac{2n\pi}{L\mu})^2
+(\frac{m}{\mu})^2)^{\frac{-s}{2}}\rfloor_{s=-1}=\]
\[
\frac{\mu^{s+1}}{2} \left(\frac{2\pi}{L}\right)^{-s}
\sum_{n=-\infty}^{\infty} (n^2
+(\frac{mL}{2\pi})^2)^{\frac{-s}{2}}\rfloor_{s=-1}\,.\]

Using equation (\ref{gam}), this can be rewritten as
\[
\frac{\mu^{s+1}}{2} \left(\frac{2\pi}{L}\right)^{-s}
\sum_{n=-\infty}^{\infty}
\frac{1}{\Gamma\left(\frac{s}{2}\right)}\int_0^{\infty}
dt\,t^{(\frac{s}{2}-1)} e^{-(n^2
+(\frac{mL}{2\pi})^2)t}\rfloor_{s=-1}\,=
\]
\[
\frac{\mu^{s+1}}{2} \left(\frac{2\pi}{L}\right)^{-s}
\frac{1}{\Gamma\left(\frac{s}{2}\right)}\int_{0}^{\infty}
dt\,t^{(\frac{s}{2}-1)}
e^{-(\frac{mL}{2\pi})^2t}\Theta(0,\frac{t}{\pi}\rfloor_{s=-1}\,,\]
where
\[\Theta(x,y)=\sum_{n=-\infty}^{\infty}e^{(2\pi xn-\pi yn^2)}\]
is a Jacobi Theta function, and has the useful inversion property
\[
\Theta(x,y)=\frac{1}{\sqrt{y}}e^{\frac{\pi
x^2}{y}}\Theta\left(\frac{x}{iy},\frac{1}{y}\right)\,.\]

Using this property, one can show that
\[
E_{\zeta}=\frac{\mu^{s+1}}{2} \left(\frac{2\pi}{L}\right)^{-s}
\frac{1}{\Gamma\left(\frac{s}{2}\right)}\left[\left(\frac{2\pi}{mL}\right)^{s-1}
\Gamma\left(\frac{s-1}{2}\right)+2\int_{0}^{\infty}
dt\,t^{(\frac{s-1}{2}-1)} e^{-(\frac{mL}{2\pi})^2
t}e^{\frac{-n^2\pi^2}{t}}\right]\rfloor_{s=-1}\,.\]

Here, you can see how the pole at $s=-1$ comes about: it comes from
the pole of the Gamma function. After performing the integral, and
developing around $s=-1$, one gets
\begin{equation} \label{Ez1}E_\zeta ^{\left( 1\right) }=-\frac m\pi
\sum\limits_{n=1}^\infty \frac 1nK_1\left( nmL\right)
+\frac{m^2L}{4\pi }\left( \left. \frac 1{s+1}\right\rfloor
_{s=-1}-\ln \left( \frac m{2\mu }\right) -\frac 12\right)\,.
\end{equation}

The series in the first term converges, due to the behaviour of the
modified Bessel function $K_1$ for $n\rightarrow\infty$.

As to the exponential cutoff regularization, it gives
\[ E_{\exp }^{\left( 1\right) }=\left. -\frac \mu
2\frac d{dt}\left( \sum_{n=-\infty }^\infty e^{-t\left( \left(
\frac{2n\pi }{L\mu }\right)  ^2+\left( \frac m\mu \right) ^2\right)
^{\frac 12}}\right)  \right\rfloor _{t=0}\,.
\]

This series can be rewritten by using the Poisson sum formula
\[
\label{sum}\sum_{n=-\infty }^\infty f\left( n\right)
=\sum_{p=-\infty }^\infty c_p\,, \] with \[
\label{cp}c_p=\int_{-\infty }^\infty dxe^{2\pi ipx}f\left(
x\right)\,.
\]

After taking the derivative, and evaluating at $t=0$, one gets
\[ E_{\exp }^{\left( 1\right) }=-\frac m\pi \sum\limits_{n=1}^\infty \frac
1nK_1\left( nmL\right) +\] \begin{equation} \label{Eexp1} \left.
\frac{m^2L}{4\pi }\left( -\ln \left( t\right)  -\ln \left( \frac m{2\mu
}\right) +2\left( \frac{mt}\mu \right)  ^{-2}-\gamma -\frac 12\right)
\right\rfloor _{t=0}\,.  \end{equation}

This result is also divergent and, as predicted, presents
divergencies in the form of poles an logarithms. The comparison of
coefficients in this expression and the one in equation (\ref{Ez1})
also shows a complete agreement with our general result.

Both results admit a renormalization, with the criterium $
E_C\rightarrow0$ for $R\rightarrow\infty$ or, equivalently, requiring
$ E_C\rightarrow0$ for $m\rightarrow\infty$ (as proposed in
reference \cite{tubo}, thus getting a physically meaningful result,
and no dependence on $\mu$.

Now, the open question is: What happens in more complicated
geometries, in particular for curved boundaries? For an answer, I
refer the reader to Professor Bordag's talk.

\bigskip

{\it Exercise 6} - Show that the $\zeta$-regularized energy for a
massive scalar field subject to periodic boundary conditions in a
$d$-dimensional box diverges for odd $d$, and is finite for even
$d$\cite{ambjorn}.

\section*{Acknowledgements}

This work was partially supported by ANPCyT (PICT'97 00039),
CONICET(PIP 0459) and UNLP (Proy. 11-X230), Argentina.

I thank C.G. Beneventano, M. De Francia and H. Falomir for many
useful comments and suggestions concerning the manuscript. I also
thank D.V. Vassilevich and the organizers of the IFSAP for making my
participation in the School possible.

A very special acknowledgement goes to Yuri and Victor Novozhilov
and to Valery Marachevsky for their warm hospitality, and to my
colleagues fighting, all around the world, for the survival of
Russian Physics.

\end{document}